\documentclass[showpacs,aps,amssymb,pra,amsmath]{revtex4}
\usepackage{float}
\usepackage{epsfig}
\usepackage{bm}
\usepackage{amsmath,graphicx}
\usepackage{subfig}
\usepackage{setspace}
\usepackage{color}

\newcommand{\beq}{\begin{equation}}
\newcommand{\eeq}{\end{equation}}
\newcommand{\bea}{\begin{eqnarray}}
\newcommand{\eea}{\end{eqnarray}}
\def\bk{{\bf k}}
\def\kv{{\bf k}}

\def\bx{{\bf x}}
\def\br{{\bf r}}
\def\rv{{\bf r}}

\begin{document}
\title{A nonlocal kinetic energy functional for an inhomogeneous two-dimensional Fermi gas}
\author{B. P. van Zyl}
\affiliation{Department of Physics, St. Francis Xavier University, Antigonish, Nova Scotia, Canada
B2G 2W5}
\author{A. Farrell}
\affiliation{Department of Physics, McGill University, Montr\'eal, QC, Canada H3A 2TB}
%\author{E. Zaremba}
%\affiliation{Department of Physics, Astronomy and Engineering Physics, Queen's University, Kingston, Ontario, Canada K7L 3N6}
\author{E. Zaremba$^{(1)}$, J. Towers$^{(2)}$, P. Pisarski$^{(1)}$,  and D. A. W. Hutchinson$^{(2),(3)}$}
\affiliation{$^{(1)}$Department of Physics, Astronomy and Engineering Physics, Queen's University, Kingston, Ontario, Canada K7L 3N6}
\affiliation{$^{(2)}$Centre for Quantum Technology, Department of Physics, University of Otago, Dunedin, New Zealand}
\affiliation{$^{(3)}$Centre for Quantum Technologies, National University of Singapore, 3 Science Drive 2, Singapore 117543}
%\author{E. Zaremba and P. Pisarski}

\date{\today}
\begin{abstract}
The average-density approximation is used to construct a nonlocal 
kinetic energy functional for an inhomogeneous two-dimensional
Fermi gas.  
%In contrast to the one and three-dimensional implementations, 
%the two-dimensional average density approximation leads to a nonlocal 
%kinetic energy functional, with no free parameters.  
This functional is then used to formulate a Thomas-Fermi von Weizs\"acker-like theory
for the description of the ground state properties of the system.
The quality of the kinetic energy functional is tested by performing a fully
self-consistent calculation for an ideal, harmonically confined, 
two-dimensional system. 
Good agreement with exact results are found, with the number
and kinetic energy densities exhibiting oscillatory structure 
associated with the nonlocality of the energy functional.
Most importantly, this functional shows a marked improvement
over the two-dimensional Thomas-Fermi von Weizs\"acker theory,
particularly in the vicinity of the classically forbidden
region. 
\end{abstract}
\maketitle
\section{Introduction}
One of the central ingredients in the density-functional theory (DFT) description~\cite{DFT} of an interacting many-body 
Fermi system is the kinetic energy (KE).  
In the Kohn-Sham (KS) scheme~\cite{KS}, the kinetic energy
functional is defined to be that of a system
of {\em noninteracting} particles having the same ground state
density as the interacting system. Formally, the KS energy
functional is,
\beq\label{KS}
E[\rho] = T[\rho] + E_{\rm int}[\rho] + \int d^3r~ v_{\rm ext}(\br)\rho(\br)~,
\eeq
where $E_{\rm int}[\rho]$ accounts for interactions
and $v_{\rm ext}(\br)$ is some externally
imposed potential. By its definition, the KS kinetic energy
functional $T[\rho]$ of $N$-independent particles is
\beq\label{KSKE}
T[\rho] = \sum_{i=1}^N \int d^3r~\phi_i^\star (\br)
\left(-\frac{\hbar^2}{2m}\nabla^2\right)\phi_i (\br)~.
\eeq
%or equivalently 
%\beq\label{KSKE2}
%T[\rho] = \sum_{i=1}^N \int d\br~\left(\frac{\hbar^2}{2m}|\nabla\phi_i(\br)|^2\right)~,
%\eeq
%and {\em not} the KE of the real interacting system.  
The $\phi_i(\br)$ are single-particle orbitals, each satisfying 
a Schr\"odingier-like equation arising from the variational minimization of Eq.~\eqref{KS}, {\it viz.,}
\beq\label{KSSE}
-\frac{\hbar^2}{2m}\nabla^2\phi_i(\br) + v_{\rm eff}(\br)\phi_i(\br) = \varepsilon_i\phi_i(\br)~,~~~~i=1,...,N
\eeq
with the effective potential given by
\beq\label{KSVeff}
v_{\rm eff}(\br) \equiv  \frac{\delta E_{\rm int}[\rho]}{\delta\rho(\br)} + v_{\rm ext}(\br)~.
\eeq
The ground state density is obtained from
\beq\label{KSden}
\rho(\br) = \sum_{i=1}^N \phi_i^\star(\br) \phi_i(\br)~,
\eeq
subject to the normalization constraint 
%finding the variational minimum of Eq.~\eqref{KS}, subject
%to the constraint of a fixed number of particles, $N$, enforced by the Lagrange multiplier, $\mu$, {\it viz.,}
%\beq\label{KS2}
% \frac{\delta T[\rho]}{\delta\rho(\br)} +  \frac{\delta E_{\rm int}[\rho]}{\delta\rho(\br)} + v_{\rm ext}(\br)-\mu=0~,
%\eeq
%with
\beq\label{norm}
N = \int d^3r~\rho(\br)~.
\eeq
%The KE and external potential energies are by far the dominant contributions to Eq.~\eqref{KS}, so ideally, one would like to
%have the exact KE functional when implementing Equation \eqref{KS2}.
%Unfortunately,  the exact noninteracting KE functional for an arbitrary inhomogeneous system is not known, so two 
%pathways for its evaluation are
%typically employed.  
%
%In the KS orbital theory, a set of  orbitals (with spin degeneracy included), $\{\phi_i(\br)\}$, are numerically
%obtained from a set of $N$ single-particle Schr\"odinger-like equations~\cite{DFT}.  The KE is then evaluated according to
%The KS KE is then an implicit functional of the density since
%
The implementation of the KS scheme
requires some approximation to be made for the generally 
unknown interaction functional, $E_{\rm int}[\rho]$.  
Although the KE is treated exactly, it suffers from
having to solve $N$ self-consistent equations, Eq.~\eqref{KSSE}, 
which can be numerically expensive.  
%Moreover, the noninteracting KE, $T[\rho]$, is dependent only implicitly on the density through
%Equation~\eqref{KSden}.

Rather than determining the KE via the KS approach, an alternative 
method is to construct an {\em explicit}
KE density functional, $T[\rho]$, thereby avoiding the
additional computational task of determining the KS orbitals.
This is the so-called orbital-free DFT, and captures the original 
spirit of DFT whereby one focuses exclusively on the density. 
Since the noninteracting KE of an arbitrary 
inhomogeneous system is not generally known as an explicit
functional of the density, approximations 
must be made for its construction.

The most primitive functional is that provided by the local-density 
approximation (LDA), in which the KE density is approximated by
that of a uniform system~\cite{DFT}. The LDA is expected
to work for systems with slow spatial variations and can be
improved by including gradient corrections which 
take spatial inhomogeneities explicitly into account. In 
three-dimensions (3D), gradient corrections can be obtained 
systematically from either a linear response~\cite{DFT,HK}, or 
a semiclassical gradient expansion approach~\cite{brack_bhaduri, 
wigner, kirkwood, kirzhnits57,jennings76}. In particular, 
the leading order functional series for the 3D noninteracting KE functional, using either approach, is given by (in what follows, a spin degeneracy factor
of two is always assumed)
\beq\label{TFvW3D}
T[\rho] = C_3 \int d^3r \rho^{5/3}(\br) + \frac{1}{9} \frac{\hbar^2}{8m} \int d^3r \frac{|\nabla \rho(\br)|^2}{\rho(\br)}~,
\eeq
where $C_3 = 3 (3\pi^2)^{2/3} \hbar^2/10m$.
The first term represents the Thomas-Fermi (TF) KE functional for a uniform system ({\it i.e.,} the LDA) while the second
term is $1/9$ the original von Weizs\"acker (vW) functional~\cite{vW}.  Equation \eqref{TFvW3D} is exact up to ${\cal O}(k^2)$ in linear response ($k/2k_F\ll 1$, where
$k_F$ is the 3D Fermi wave number), but works quite well even for strongly inhomogeneous systems.
Higher order terms involving gradients of the
density may be added to Eq.~\eqref{TFvW3D},  but such terms are not guaranteed to improve the quality of the KE
functional, and in some circumstances, may actually diverge~\cite{DFT,hodges73}.  

In applications, the functional
\beq\label{vW3D}
T[\rho] = C_3 \int d^3r \rho^{5/3}(\br) + \lambda_{\rm vW} \frac{\hbar^2}{8m} \int d^3r \frac{|\nabla \rho(\br)|^2}{\rho(\br)}~
\eeq
is often used,
where $\lambda_{\rm vW}$ (the vW coefficient) is an adjustable 
parameter.
The total energy in this approximation is then given by the Thomas-Fermi von Weizs\"acker (TFvW) functional
\beq\label{I4}
E[\rho] = C_3 \int d^3r \rho^{5/3}(\br) + \lambda_{\rm vW} \frac{\hbar^2}{8m} \int d^3r \frac{|\nabla \rho(\br)|^2}{\rho(\br)} +E_{\rm int}[\rho] + \int d^3r ~v_{\rm ext}(\br)\rho(\br)~.
\eeq
The parameter $\lambda_{\rm vW}$ can be tuned to obtain the best 
agreement with the energy generated by a KS orbital calculation.  

The minimization of Eq.~\eqref{I4} with respect to the density,
$\rho(\br)$, for a fixed number
of particles, leads to a single Schr\"odinger-like equation for the vW wave function, $\psi(\br) \equiv \sqrt{\rho(\br)}$,
\beq\label{SEvW3D}
-\lambda_{\rm vW}\frac{\hbar^2}{2m} \nabla^2\psi(\br) + v_{\rm eff}(\br) \psi(\br) = \mu \psi(\br)
\eeq
where the effective one-body potential is given by
\beq\label{veff3D}
v_{\rm eff}(\br) = \frac{5}{3} C_3 \rho^{2/3}(\br) +  \frac{\delta E_{\rm int}[\rho]}{\delta\rho(\br)} + v_{\rm ext}(\br)~.
\eeq
The vW term, $-(\lambda_{\rm vW}\hbar^2/2m) \nabla^2\psi(\br)$, 
is known to provide a smooth decay of the spatial density into 
the classically forbidden region, and cures the unphysical, 
sharp cutoff of the density at the classical turning point found 
within the TF approximation~\cite{brack_bhaduri}. This approach
is easy to implement and computationally inexpensive ({\it i.e.,} 
only a single self-consistent equation, Eq.~\eqref{SEvW3D}, needs 
to be solved). For 3D Coulombic systems with local exchange, it
is referred to as the Thomas-Fermi-Dirac-von Weisz\"acker (TFDW)
theory and has yielded good results in various 
applications~\cite{chizmeshya88,zaremba_tso}.

%When $E_{\rm int}[\rho]$ is the interaction functional associated with the Coulomb potential, and evaluated
%according to the Dirac decomposition of the two-body density matrix~\cite{dirac},  Eq.~\eqref{I4} is known as the 
%Thomas-Fermi-Dirac-von Weizs\"acker (TFDW) approximation, which has
%been highly successful in the DFT of degenerate 3D electron gases~\cite{zaremba_tso,zaremba96,chizmeshya88}.  The 3D TFDW is exceedingly easy to implement, 
%computationally inexpensive ({\it i.e.,} only a single self-consistent equation, Eq.~\eqref{SEvW3D}, needs to be evaluated), and has the
%virtue of being able to completely dispense with additional boundary conditions, which are usually specified by considering the flux of conserved
%quantities ({\it e.g.,} mass, momentum) at the boundary of the system.

For the inhomogeneous two-dimensional (2D) Fermi gas, it is natural to try to follow the same formulation as  the 3D TFvW theory outlined above.  Unfortunately, 
in 2D, it is known that neither the linear response~\cite{HK}, or 
semiclassical 
methods~\cite{geldhart86,holas91,salasnich,koivisto,PhD,putaja,vanzyl11} yield any gradient corrections whatsoever.  This is troublesome, since we know that
the LDA cannot be exact for an inhomogenous system.  
Nevertheless, gradient corrections introduced in an {\it ad hoc}
fashion do provide a more realistic description of 2D
density distributions. Specifically, the 2D analogue of
Eq.~\eqref{vW3D} reads
\beq\label{vW2D}
T[\rho] = C_2 \int d^2r ~\rho^{2}(\br) + \lambda_{\rm vW} \frac{\hbar^2}{8m} \int d^2r \frac{|\nabla \rho(\br)|^2}{\rho(\br)}~,
\eeq
where $C_2 = \pi\hbar^2/2m$.  Equation \eqref{vW2D} has been
used to construct the 2D version of the TFDW theory for an 
inhomogeneous two-dimensional electron gas 
(2DEG)~\cite{zaremba_tso,PhD}, and 
more recently, to describe the equilibrium properties of a 2D
harmonically trapped, spin polarized dipolar Fermi
gas~\cite{vanzyl_pisarski}. Although the vW term cannot be
justified on the basis of a gradient expansion, the 2D TFDW
theory was nonetheless successful in various applications
to 2DEGs~\cite{vanzyl1,vanzyl2,vanzyl3,vanzyl4}.
%This is not so much a condemnation of the  theory, but rather is meant to highlight
%that standard techniques for attributing a nonlocal character to the KE functional based on gradient corrections are not applicable in two-dimensions.  
%develope a KE functional rooted in a systematic treatment of beyond LDA physics in confined Fermi systems. 
%have an approach for constructing the KE functional of an inhomogeneous system that is not sensitive to the dimensionality. 
%In fact, one of our objectives is to maintain the exceedingly simple (yet effective) mathematical formulation of the TFvW
%theory, while developing a KE functional rooted in a systematic treatment of beyond LDA physics in confined Fermi systems. 

In this paper, we
make use of the average-density approximation (ADA)
to define a nonlocal KE functional, which allows us to systematically treat a spatially inhomogeneous 2D Fermi gas
without the use of any {\em ad hoc}, {\it e.g.,} vW, gradient corrections.  This functional is then utilized
to formulate a self-consistent TFvW-like theory for the 2D 
inhomogeneous Fermi gas. The efficacy of the theory is tested 
in Sec.~III by comparing our
self-consistent calculations for the ground state properties of an ideal, harmonically trapped Fermi gas, 
with exact results, and with the results of  the 2D TFvW theory using an optimal vW coefficient ({\it i.e.,} the value of $\lambda_{\rm vW}$ for which Eq.~\eqref{vW2D}  
yields the exact KE for the TFvW self-consistent density)~\cite{vanzyl_pisarski}.
In Sec.~IV, we present our closing remarks.

%%%%%%%%%%%  ADA %%%%%%%%%%%%%%%
\section{The average-density approximation}\label{ADA}
The ADA was first proposed in the late $1970$s by Alonso {\it
et al.}~\cite{alonso} and Gunnarsson {\it et al.}~\cite{gunnarsson}, 
as a way to go beyond the LDA for calculations of the
exchange and correlation energies of non-uniform electron
systems.  These same ideas were later applied in the construction 
of nonlocal kinetic energy
functionals in 1D~\cite{herring,comb97,gonzalez98} and 3D~\cite{chacon85,garcia96,wang,chai2007,chacon07,perrot} systems.  It is within the latter context that we wish to briefly review the essential ideas behind the ADA.
In this section of the  paper, we will use atomic units ($\hbar=m=1$).

At the heart of  the ADA is the specification of the nonlocal KE functional in terms of an average-density (analogous expressions hold for 1D and 2D), {\it viz.,}
\beq\label{W1}
T_{\rm nl}[\rho(\br)] = \int d^3r~\rho(\br) t(\bar{\rho}(\br))~,
\eeq
where $t(x)$ is the kinetic energy per particle of the {\em uniform} system, and the average density, $\bar{\rho}(\br)$, is defined as
\beq\label{W2}
\bar{\rho}(\br) = \int d^3r' \rho(\rv') w(\rv'-\rv; \rho(\rv)) ~.
\eeq
The nonlocal character of the inhomogeneous system is then captured by the weight function $w(\br;\rho)$, which is normalized according to
\beq\label{W3}
\int d^3x~ w(\bx;\rho) = 1~.
\eeq
The normalization ensures that  Eq.~\eqref{W1} reproduces the exact 
kinetic energy of a uniform system.  The weight function itself 
is fully specified by also
demanding that the second functional derivative of Eq.~\eqref{W1}, when evaluated for a uniform system of density, $\rho_0$, leads to the
{\em exact} static linear response function~\cite{HK}, {\it viz.,}
\beq  \label{W4}
\mathcal{F}\left[\frac{\delta^2 T_{\rm nl}[\rho]}{\delta \rho(\rv) \delta \rho(\rv')} \right]_{\rho(\br)=\rho_0}= -\frac{1}{\chi_0(\bk)}~,
\eeq
where $\mathcal{F}$ denotes a Fourier transform (FT) from $\rv-\rv'$ to $\kv$. Performing the functional derivatives is straightforward, and after
taking the FT, we obtain
the following general result~\cite{chacon85}
\beq\label{W5}
\mathcal{F}\left[ \frac{\delta^2 T_{\rm nl}[\rho] }{\delta
\rho(\br) \delta \rho(\br') }\right]_{\rho(\br)=\rho_0}
=2t'(\rho_0) w(k; \rho_0)+ \rho_0 t''(\rho_0)w(k; \rho_0)^2+
2\rho_0 t'(\rho_0) \frac{d w(k; \rho_0)}{d\rho_0}~,
\eeq
where $ w(k; \rho_0)$ is the FT of $w(|\rv-\rv'|; \rho_0)$.
In principle, once the weight function is known, Eq.~\eqref{W1} defines a nonlocal KE functional, which will exactly reproduce the KE of the uniform gas, 
and whose second functional
derivative is {\em exact} in the uniform limit~\cite{note_nl}. 
Whether this prescription for the KE functional also provides 
a good description for an inhomogeneous system must be established 
separately.

In practice, however, the right-hand side of Eq.~\eqref{W4} is divergent in the 
$k \to \infty$ limit, and this divergence occurs regardless of
dimensionality.  This is an undesirable feature, as the 
weight function will also inherit the divergence.  One way
to avoid this issue is to remove the offending terms from the right-hand side of Eq.~\eqref{W4}, 
and then solve for a {\em new} weight function with no divergent behaviour.
The details of this procedure requires an explicit analytical form for $\chi_0(k)$, so it is necessary to take up the rest of our discussion
in strictly 2D.

%%%%%%%%%%%%%%%% TWO-DIMENSIONAL NONLOCAL KE FUNCTIONAL %%%%%%%%%%%%%%%%%%%
\subsection{Two-dimensional nonlocal kinetic energy functional}\label{2DnlKE}
The exact static response function, $\chi_0(k)$, for a uniform  2D Fermi gas of density $\rho_0$ is given by~\cite{stern} 
\beq\label{NL1}
\chi_0(\eta) = \begin{cases}
-\dfrac{1}{\pi},&\eta <1 \\
-\dfrac{1}{\pi}\left(1-\sqrt{1-\dfrac{1}{\eta^2}}\right),&\eta \geq1~,
\end{cases}
\eeq
with $\eta\equiv k/(2k_F)$, and $k_F=\sqrt{2\pi\rho_0}$. This
expression is the 2D analogue of the 3D Lindhard
function~\cite{lindhard}.

It is useful to write Eq.~\eqref{NL1} as  
\beq\label{NL2}
-\frac{1}{\chi_0(\eta)} =\pi F(\eta)~, 
\eeq
where
\beq\label{NL3}
F(\eta) = \begin{cases}
1,&\eta <1~,\\
\dfrac{1}{1-\sqrt{1-\dfrac{1}{\eta^2}}},&\eta \geq1~.
\end{cases}
\eeq
Note that $F(\eta)$ has a pole at $\eta \to \infty$, and is 
piecewise continuous at $\eta=1$.
Equation \eqref{W4} then amounts to
\beq  \label{NL4}
2t'(\rho_0) w(k; \rho_0)+ \rho_0 t''(\rho_0)w(k; \rho_0)^2+ 
2\rho_0 t'(\rho_0) \frac{d w(k; \rho_0)}{d\rho_0}
= \pi F(\eta)~.
\eeq
%{\bf The following sentence is obscure!}
%It is now clear why the 2D TFvW KE functional, Eq.~\eqref{vW2D}, is formally unjustified; the 2D TFvW KE functional in the uniform limit {\em does not} 
%agree with the linear response theory
%beyond $k=0$, since $F(k/2k_F)$ is a constant for $k/2k_F < 1$~\cite{note_nl}.

The kinetic energy per particle for the uniform 2D gas is given by
\beq\label{NL5}
t(\rho_0) = \frac{\pi}{2}\rho_0~,
\eeq
and we see that Eq.~\eqref{NL4}  reduces to a first-order ordinary differential equation (ODE).
In view
of the dependence of the response function on the scaled wavevector 
$\eta$, we will assume that the weight function has a similar 
dependence, namely $w(k;\rho_0) \equiv w(k/2k_F) = w(\eta)$.    
As a result,
\beq
 \rho_0 \frac{d w(k; \rho_0)}{d\rho_0} = -\frac{\eta}{2}
w'(\eta)~,
\eeq
and the ODE for the weight function then reads
\beq\label{NL6}
w(\eta)-\frac{\eta}{2}w'(\eta)  =  F(\eta)~.
\eeq
The normalization condition, Eq.~\eqref{W3}, in real space
implies that $w(\eta=0)=1$, which is automatically satisfied by
the solution of Eq.~(\ref{NL6}). 
Once the solution to Eq.~\eqref{NL6} is obtained, the nonlocal 
KE functional is given by
\beq\label{nlke1}
T_{\rm nl}[\rho] =  \frac{\pi}{2}\int d^2r \int d^2r'~\rho(\br'){w}(\br-\br';\rho(\br)){\rho}(\br)~.
\eeq

It is straightforward to show that $w(\eta)$ is given by 
\beq\label{sol1}
w(\eta) = -2 \eta^2 \int \frac{F(\eta)}{\eta^3}d\eta + c \eta^2~,
\eeq
where $c\eta^2$ is the solution of the homogeneous equation.  For $\eta <1$, $F(\eta)=1$, and we obtain
\beq\label{sol2}
w(\eta) = 1+c_1\eta^2~,
\eeq
where $c_1$ is fixed by demanding continuity of $w(\eta)$ at $\eta=1$.  
For $\eta \geq 1$, we have
\bea\label{sol3}
w(\eta) &=& -2\eta^2 \int_1^{\eta} \frac{1}{t^3} \frac{1}
{1-\sqrt{1-1/t^2}}dt +c_2 \eta^2\nonumber \\
&=& -4 \eta^2\ln \eta+2\eta\sqrt{\eta^2-1} -2\eta^2 \ln\left[1+
\sqrt{1-\frac{1}{\eta^2}}\right] + c_2\eta^2~.
\eea
In the $\eta\to\infty$ limit, Eq.~\eqref{sol3} has the following behaviour
\beq\label{sol4a}
w(\eta) \sim -4 \eta^2\ln\eta + 2\eta^2
-\ln 4\eta^2+c_2\eta^2-\frac{1}{2}\cdots~.
\eeq
The $\eta^2$ divergence in Eq.~\eqref{sol4a} can be eliminated 
by choosing $c_2=\ln 4-2$, but as
discussed above, $w(\eta)$ still has a rather nasty divergence
$-4 \eta^2\ln\eta$ for $\eta\to\infty$.  The origin of this divergence can be traced back to the $\eta\to\infty$ behaviour of $F(\eta)$, namely
\beq\label{sol5}
F(\eta) \sim 2 \eta^2~.
\eeq
Therefore, 
deleting the  $\eta\to\infty$ terms from $F(\eta)$ will yield a new weight function with no divergent behaviour, {\it viz.,}
\beq\label{weight1}
 w_0(\eta)-\frac{\eta}{2}w_0'(\eta)  =  F(\eta)-2 \eta^2 ~.
 \eeq
Note that here, $w_0(0)=1$, so the normalization of the weight
function is preserved.  The solution to Eq.~\eqref{weight1} is given by
\begin{eqnarray}\label{weight2}
w_0(\eta) &=& \left[ {4\eta^2\ln\eta}+1+\left(\ln 4-3\right)\eta^2\right]\Theta(1-\eta) \nonumber\\ 
&+&\left[2\eta\sqrt{\eta^2-1} + \eta^2(\ln 4-2)  
-2\eta^2 \ln\left( 1+\sqrt{1-\frac{1}{\eta^2}}\right)\right]\Theta(\eta-1)~,
\end{eqnarray} 
which no longer exhibits any divergences. 

In order to satisfy Eq.~\eqref{NL4}, however, we must now write 
the nonlocal KE functional as
\beq\label{NLextra3}
T_{\rm nl}[\rho] =  \frac{\pi}{2}\int d^2r \int d^2r'~\rho(\br'){w_0}(\br-\br';\rho(\br)){\rho}(\br)+T_{\rm vW}[\rho]~,
\eeq
where
\beq\label{nlke2}
T_{\rm vW}[\rho] = \frac{1}{8}\int d^2r \frac{|\nabla \rho(\br)|^2}{\rho(\br)}~,
\eeq
is the vW functional satisfying
\beq
\mathcal{F}\left[ \frac{\delta^2 T_{\rm vW}[\rho] }{\delta
\rho(\br) \delta \rho(\br') }\right] _{\rho(\br)=\rho_0} =2\pi
\eta^2~.
\eeq
The addition of the vW functional in Eq.~(\ref{NLextra3})
ensures that Eq.~\eqref{NL4} holds with a weight function which
is the solution of Eq.~(\ref{weight1}).

\begin{figure}[H]
\centering \scalebox{0.27}
{\includegraphics{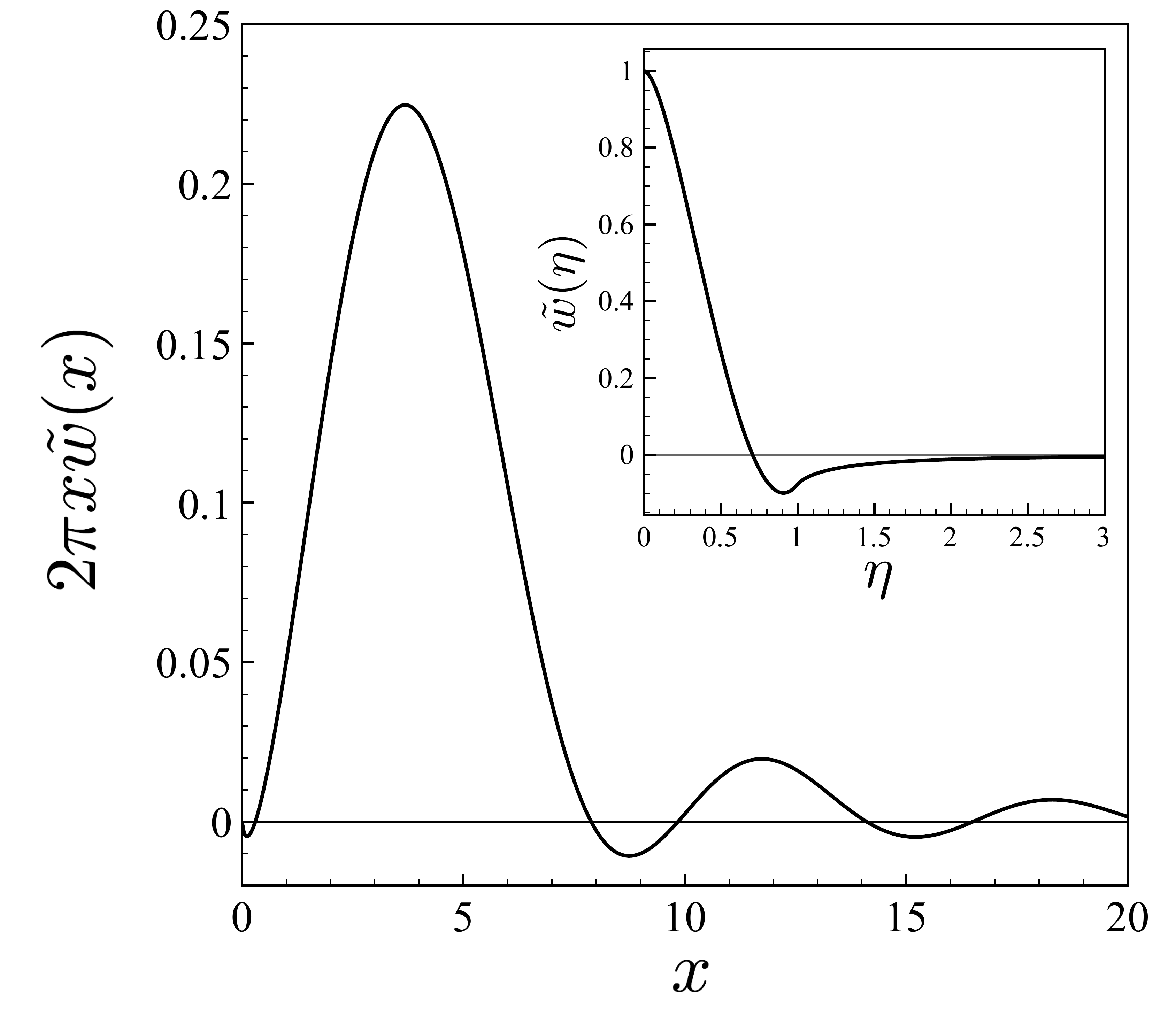}}
\caption{The main figure presents the 2D real-space weight function, while the figure inset displays its  Fourier transform.  Note that
we have dropped the subscript on $\tilde{w}$, as discussed in Sec.~\ref{TFvW-like} of the text.  Here, 
$x \equiv 2 k_F r$ is the dimensionless spatial coordinate.}
\label{fig0}
\end{figure}

The fact that $w_0(\eta)$ has the limiting value of $-1/2$ for 
$\eta\to\infty$ implies that the weight function has a 
$\delta(\br)$ contribution in real space. One can remove the 
Dirac-delta contribution by simply defining
yet another weight function
\beq\label{weight3}
w_{\frac{1}{2}}(\eta) \equiv w_0(\eta)+\frac{1}{2}~.
\eeq
In terms of this weight function, the nonlocal KE functional in
Eq.~(\ref{NLextra3}) is given by
\beq\label{NL13}
T_{\rm nl}[\rho] =  \frac{\pi}{2}\int d^2r \int d^2r'~\rho(\br'){w}_\frac{1}{2}(\br-\br';\rho(\br)){\rho}(\br) - \frac{1}{2}T_{\rm TF}[\rho] + T_{\rm vW}[\rho]~,
\eeq
where
\beq\label{weight4}
T_{\rm TF}[\rho] = \frac{\pi}{2} \int d^2r \rho^2(\br)~, 
\eeq
and
\beq\label{weight5}
\mathcal{F}\left[ \frac{\delta^2 T_{\rm TF}[\rho] }{\delta \rho(\br) \delta \rho(\br') }\right] _{\rho(\br)=\rho_0} =\pi~.
\eeq
Since this new weight function has the limiting value 
${w}_\frac{1}{2}(0)=3/2$, Eq.~(\ref{NL13}) can be written
alternatively as
\bea\label{weight6}
T_{\rm nl}[\rho] &=&  \frac{3\pi}{4}\int d^2r \int d^2r'~\rho(\br')\tilde{w}_\frac{1}{2}(\br-\br';\rho(\br)){\rho}(\br) - \frac{1}{2}T_{\rm TF}[\rho] + T_{\rm vW}[\rho]\nonumber \\
&\equiv& \frac{3}{2} T_{\rm ADA}[\rho;\tilde{w}_\frac{1}{2}] - \frac{1}{2}T_{\rm TF}[\rho] + T_{\rm vW}[\rho]~,
\eea
where $\tilde{w}_\frac{1}{2}(\eta) \equiv \frac{2}{3}
{w}_\frac{1}{2}(\eta)$ now satisfies the normalization
condition $\tilde{w}_\frac{1}{2}(0)=1$.  In Fig.~\ref{fig0}, we plot the weight function, $\tilde{w}_{\frac{1}{2}}$ in both
real space (main figure) and in Fourier space (figure inset).  It is interesting to note that the 2D real-space weight function is qualitatively similar to the one found in 3D~\cite{garcia96},
although in 3D, an analytical solution for the weight function is not possible.

Before proceeding any further
a few comments on Eq.~\eqref{weight6} are in order. First, we note that the second term in Eq.~\eqref{weight6} has exactly the same form as the TF KE
functional, but
with a {\it negative} coefficient.  Moreover,
the last term in Eq.~\eqref{weight6} is exactly the vW functional, which has naturally arisen in our formulation
owing to the requirement that the second functional derivative of $T_{\rm nl}[\rho]$ agree with the homogeneous response function for all wave vectors.  In view of the forms of 
the second and third terms in Eq.~\eqref{weight6}, we adopt the term TFvW-like to describe our DFT for the inhomogenous 2D Fermi gas.   

It should be noted the vW KE {\em density} is only
defined up to the addition of a function which integrates to zero 
over all space.  In particular, using the
definition $\psi(\br) = \sqrt{\rho(\br)}$, it can easily be shown that
\bea\label{C1}
\frac{1}{8}\frac{|\nabla\rho(\br)|^2}{\rho(\br)} &=&\frac{1}{2}|\nabla\psi(\br)|^2 \nonumber \\
&=&\frac{1}{4}\nabla^2\rho(\br) -\frac{1}{2}\sqrt{\rho(\br)}\nabla^2\sqrt{\rho(\br)}~.
\eea
The term $\nabla^2\rho(\br)/4$ in Eq.~\eqref{C1} is the divergence of a vector field which vanishes at infinity; by the divergence theorem, it will not contribute
to the kinetic energy, and can therefore be dropped.
It follows that the vW KE density may be expressed alternatively as
\beq\label{C3}
\tau_{\rm vW}(\br) = \frac{1}{8}\frac{|\nabla\rho(\br)|^2}{\rho(\br)}~,
\eeq
or
\bea\label{C4}
\tau^{(1)}_{\rm vW}(\br) 
&=&-\frac{1}{2}\sqrt{\rho(\br)}\nabla^2\sqrt{\rho(\br)}~,
\eea
both of which integrate to the same vW kinetic energy. 
As a result, one may consider various forms for the nonlocal KE density without changing the 
physical properties of the system, as they all lead to the same nonlocal KE functional.

While Eq.~\eqref{weight6} is exact for a uniform system, and 
exactly reproduces the homogeneous gas linear response function,
it does not yield the exact KE for a localized single-particle 
state which
is captured by the vW functional itself. It would be ideal if 
the nonlocal functional could also produce this limit,
thereby acting as a ``bridge'' between the uniform and vW
limits. One can try to achieve this objective by 
considering a generalized functional of the form~\cite{chacon85}
\beq\label{G1}
T_{\rm nl}[\rho] = (1+\alpha)T_{\rm ADA}[\rho;\tilde{w}_\alpha]-\alpha T_{\rm TF}[\rho]+ T_{\rm vW}[\rho]~.
\eeq
Here, $\alpha$ is ostensibly a free, adjustable
parameter which is to be chosen such that Eq.~\eqref{G1} is 
exact in both the uniform and vW limits.
Clearly, putting $\alpha=1/2$ recovers Equation~\eqref{weight6}.  
Following exactly the same analysis as above, we have
\beq\label{wtildealpha}
\tilde{w}_\alpha(\eta) = \frac{w_0(\eta)}{1+\alpha} + \frac{\alpha}{1+\alpha}~.
\eeq
Inserting Eq.~\eqref{wtildealpha} into Eq.~\eqref{G1}
immediately leads to Eq.~\eqref{weight6}.  Therefore, in 2D, $\alpha$ is {\em not} a free parameter, since any value of $\alpha$ leads
to the same nonlocal KE functional, {\it viz.,} 
Eq.~\eqref{weight6} (or equivalently, Eq.~(\ref{NLextra3})). 
This means that it is {\em not possible} to have the ADA nonlocal 
KE functional yield the exact KE in the vW limit.
This is in stark contrast to the  3D 
case where $\alpha$ is a tuneable parameter~\cite{chacon85}.  
Specifically, in 3D, a special value of $\alpha^{\rm 3D} \simeq 2/9$  
ensures that the nonlocal KE functional is exact in the uniform and 
vW limits. The 1D geometry is similar to the 3D situation, in that 
the KE functional depends on the the value chosen for 
$\alpha$~\cite{comb97}.
The fundamental difference between 2D and 1D/3D is that the ODE for 
the weight function is {\em linear} in the former, but nonlinear
in the latter.  It is the linearity of the weight function ODE which ultimately accounts for the $\alpha$-independence in two-dimensions.  

For an inhomogeneous system, we continue to use the form of the 
weight function obtained from the uniform system, but now with
the wave vector scaled by the local Fermi wave vector
$k_F(\br)=\sqrt{2\pi\rho(\br)}$.  This of course is an 
approximation whose validity must be verified separately 
(see Sec.~III below).
It is noteworthy that the 2D geometry has allowed for an exact, analytical expression for the weight function.  In contrast, the 1D and 3D
cases require a numerical evaluation of a non-linear first-order ODE for the weight function, which adds an additional layer of complexity to the 
computational implementation of the ADA  KE functional.  
%%%%%%%%%%%%%%%%% TFVW THEORY IN TWO-DIMENSIONS %%%%%%%%%%%%%%%%%%%%%%%%%%
\subsection{TFvW-like theory in two-dimensions}\label{TFvW-like}
We are now in a position to present the TFvW-like theory 
based on the ADA for the nonlocal KE functional, Eq.~\eqref{weight6}, which is given by
\bea\label{SC1}
E[\rho] &=& \frac{3}{2}\int d^2r \int d^2r' \frac{\pi}{2} \rho(\br')\tilde{w}(\br-\br';\rho(\br))\rho(\br) 
-\frac{1}{2} \int d^2r \frac{\pi}{2}\rho(\br)^2 + \frac{1}{8} \int d^2r \frac{|\nabla \rho(\br)|^2}{\rho(\br)}\nonumber \\
&+& E_{\rm int}[\rho(\br)] +  \int d^2r~v_{\rm ext}(\br) \rho(\br)~.
\eea
Henceforth, we drop for convenience the $1/2$ subscript on 
$\tilde{w}(\eta)$.
The variational minimization of Eq.~\eqref{SC1} for a fixed number of particles yields the defining equations for the TFvW-like theory, {\it viz.,}
\beq\label{SC8}
-\frac{1}{2} \nabla^2\psi(\br) + v_{\rm eff}(\br) \psi(\br) = \mu \psi(\br)~,
\eeq
where
\beq\label{SC9}
v_{\rm eff}(\br) = -\frac{\pi}{2} \psi(\br)^2 + \phi(\br) + v_{\rm ext}(\br) + \frac{\delta E_{\rm int}}{\delta \rho(\br)}~,
\eeq
\bea\label{SC10}
\phi(\br)=\frac{3\pi}{4} \int
\frac{d^2k}{(2\pi)^2}\int d^2 r_1
e^{i\kv\cdot(\rv-\rv_1)}\left[\Omega
\left(\frac{k}{2k_F(\rv)}\right)+ \tilde{w}
\left(\frac{k}{2k_F(\rv_1)}\right)\right] \rho(\rv_1)~,
\eea
\beq\label{Omega}
\Omega \left(\eta \right)\equiv \frac{2}{3} \left(F(\eta)+\frac{1}{2}-2 \eta^2\right)~.
\eeq
and 
\beq\label{SC11}
N(\mu) = \int d^2r~|\psi(\br)|^2~.
\eeq
Note that Eqs.~\eqref{SC8}--\eqref{SC11} do not actually require
the evaluation of $\tilde{w}(\br;\rho)$ in real-space, which allows us to fully exploit the
analytical expression for the weight function.  We observe that
the first term in the square braces of Eq.~\eqref{SC10} 
depends locally on the density whereas 
the second term, involving $\tilde{w}$, has a nonlocal
dependence.

We have clearly reached our objective of preserving the simple
mathematical framework of the TFvW theory, without having to 
introduce any {\em ad hoc} gradient terms to the KE functional.  
%{\bf I don't understand the following sentence \{}Indeed, 
%Eq.~\eqref{SC8} can be made formally identical to the original 
%2D TFvW theory~\cite{PhD} by simply replacing 
%$\nabla^2/2$ by $\lambda_{\rm vW}\nabla^2/2$.{\bf \}} 
It should also be noted 
that as in any orbital-free DFT scheme, {\it e.g.,} the TFvW theory, the computational expense 
of the TFvW-like theory does not scale with the number of particles.
In Sec.~\ref{sc} we will investigate the self-consistent 
solutions to the TFvW-like theory in detail.  

%%%%%%%%%%%%%%   APPLICATION %%%%%%%%%%%%%%%%%%%%%%%%%%%%%%%%%%%%%%%%%%%%%
\section{Application: Harmonically trapped Fermi gas}\label{app}
\subsection{Tests using exact densities}
As a first step in determining the quality of our nonlocal KE 
functional, we will utilize exact results available for an
ideal Fermi gas in a 2D isotropic harmonic oscillator (HO) potential,
\beq\label{HO}
v_{\rm ext}(\br)= \frac{1}{2} {m\omega^2 r^2}~.
\eeq
Henceforth, all lengths are scaled by the
HO oscillator length, $\ell_{\rm osc} = \sqrt{\hbar/m\omega}$,
and energies by the HO energy, $\hbar\omega$.  

\begin{figure}[t]
\centering \scalebox{0.8}
{\includegraphics{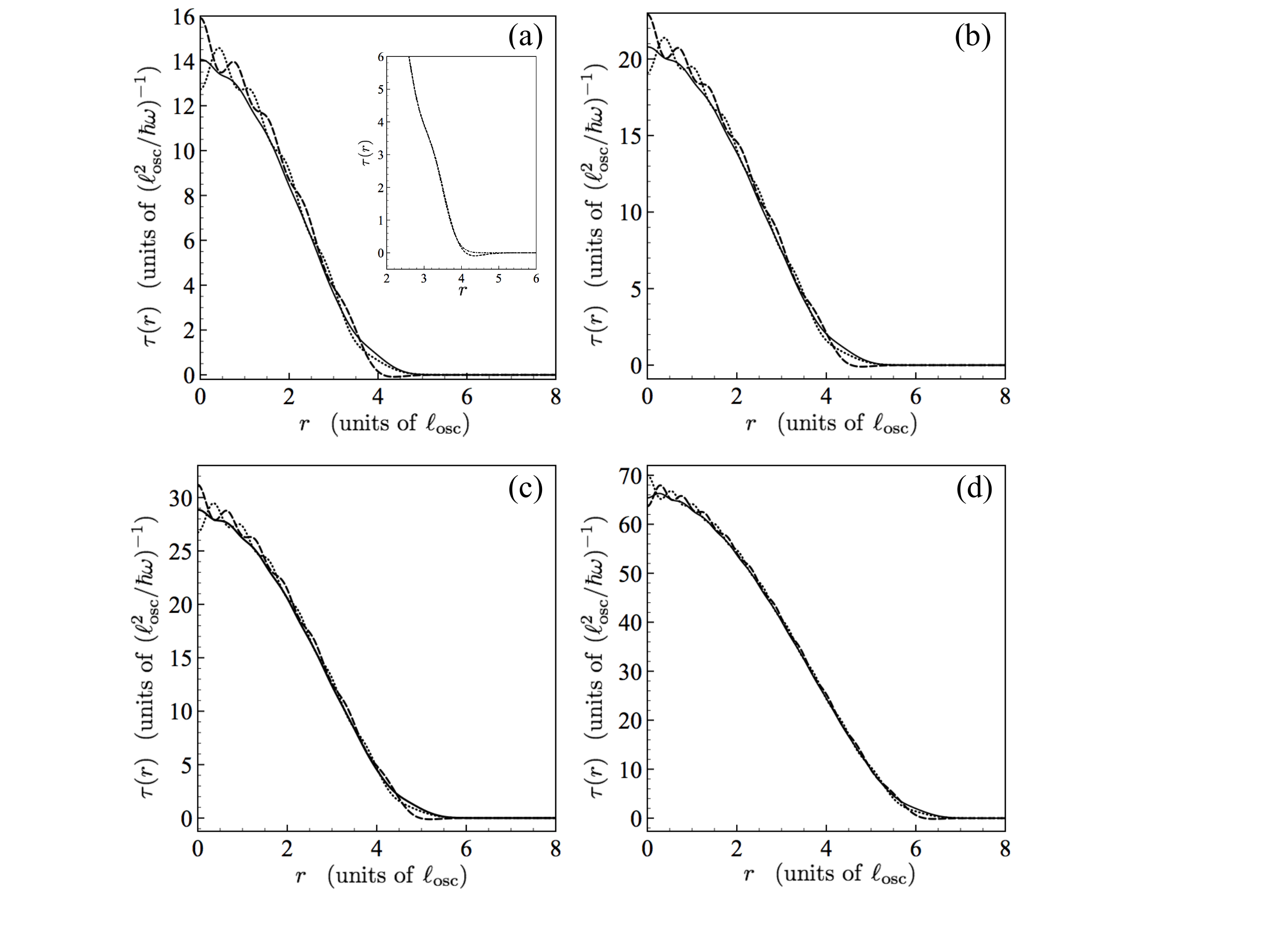}}
\caption{The kinetic energy densities, $\tau_{\rm ex}(r)$ 
(dotted curves), $\tau^{(1)}_{\rm ex}(r)$ (dashed curves) and 
$\tau_{\rm nl}(r)=\tau_{\rm nl}(\rho_{\rm ex}(r))$  (solid curves), 
for  (a) $N=90$, (b) $N=132$, (c) $N=182$, (d) $N=420$, particles.  
The inset to panel (a) shows $\tau^{(1)}_{\rm ex}(r)$ (dashed curve) 
and $\tau_{\rm TF}(\rho_{\rm ex}(r))$ (dot-dashed curve) near 
the tail region.
The axes of the figure inset are scaled as in the main figure.}
\label{fig1}
\end{figure}

For an arbitrary number of closed shells, the exact 2D spatial 
density, and its FT, are given respectively 
by~\cite{brack_vanzyl,shea_vanzyl} 
\beq\label{A2}
\rho_{\rm ex} (r) = \frac{2}{\pi} \sum_{n=0}^M (-1)^n (M-n+1) L_n(2r^2)e^{-r^2}~,
\eeq
and
\beq\label{FTden}
\rho_{\rm ex} (k) = 2 L_M^2(k^2/2) e^{-k^2/4}.
\eeq
In the above, the shell index $M$ defines the number of filled
shells, $M+1$, and $L_n^\beta (x)$ is an associated Laguerre polynomial~\cite{grad}.  
The total particle number, $N$, is given in terms of $M$ by
\beq
N(M) = (M+1)(M+2)~.
\eeq
The exact KE density for the 2D HO is also known, and may be written in three different forms~\cite{brack_vanzyl,shea_vanzyl}:
\bea\label{keden}
\tau_{\rm ex}(r) &=&  \sum_{\epsilon_k\le\epsilon_F} |\nabla \phi_k (\br)|^2 \nonumber \\
&=&\frac{1}{\pi} \sum_{n=0}^M (-1)^n (M-n+1) (M-3n+r^2) L_n(2r^2)e^{-r^2}~,
\eea
\bea\label{keden1}
\tau^{(1)}_{\rm ex}(r) &=&  -\sum_{\epsilon_k\le\epsilon_F}\phi_k(\br)\nabla^2\phi_k(\br) \nonumber \\
&=& \frac{1}{\pi} \sum_{n=0}^M (-1)^n (M-n+1) (M+n+2-r^2) L_n(2r^2)e^{-r^2}~,
\eea
and
\beq\label{keden2}
\tau^{(2)}_{\rm ex}(r) = \frac{\tau_{\rm ex}(r)+\tau^{(1)}_{\rm ex}(r)}{2} = \frac{1}{\pi} \sum_{n=0}^M (-1)^n (M-n+1)^2 L_n(2r^2)e^{-r^2}~.
\eeq
In the above, $\phi_k(\br)$ are the orthonormal HO eigenstates 
and $\epsilon_F = M+1$.  Recall that we have included a spin factor of two in the KE densities defined
in Equations~\eqref{keden} and \eqref{keden1}.
When integrated over space, all three KE densities give the same 
exact KE, 
\beq\label{A3}
T_{\rm ex} = \frac{N}{6}\sqrt{1+4N}~.
\eeq
Although $\tau_{\rm ex}(r) \geq 0$ and $\tau^{(2)}_{\rm ex}(r) 
\geq 0$ for all $r$,  $\tau^{(1)}_{\rm ex}(r)$ takes on small, 
negative values in the tail region (see 
Fig.~\ref{fig1})~\cite{brack_vanzyl}.  
This behaviour can be understood by noting that the KE energy
densities behave as~\cite{howard2010}
\beq\label{keden_large}
\tau_{\rm ex}(r)  \sim %\underset{r\to\infty}{\longrightarrow}
\frac{1}{8}\frac{|\nabla\rho_{\rm ex}(r)|^2}{\rho_{\rm ex}(r)}
\geq 0~,
\eeq
and
\beq\label{keden1_large}
\tau^{(1)}_{\rm ex}(r) \sim %\underset{r\to\infty}{\longrightarrow}  
\frac{1}{2}\sqrt{\rho_{\rm ex}(r)}(-\nabla^2)
\sqrt{\rho_{\rm ex}(r)} \leq 0~,
\eeq
in the classically forbidden region $r \gg R_{\rm TF}$, where
$R_{\rm TF}= \sqrt{2}N^{1/4}$ is the TF radius. We thus see that
$\tau_{\rm ex}(r)$ asymptotically approaches the form of the vW KE 
density in Eq.~(\ref{C3})
while $\tau^{(1)}_{\rm ex}(r)$ approaches the form in
Eq.~(\ref{C4}). 
In addition, $\tau_{\rm ex}(r)$ and  $\tau^{(1)}_{\rm ex}(r)$
have oscillations associated with shell structure. These
oscillations are exactly out of phase (see Fig.~\ref{fig1}) and
results in $\tau^{(2)}_{\rm ex}(r)$ in Eq.~\eqref{keden2}
being a smooth function~\cite{bhaduri70}.

One way of investigating the quality of the KE functional in
Eq.~(\ref{weight6}) is to see what it yields for the KE when
the exact density is inserted, {\it i.e.,} using Eq.~\eqref{A2} in~\eqref{weight6} and integrating
over all space.
In view of the circular symmetry of the harmonically-confined
system being considered, the nonlocal KE functional can be
written as
\beq\label{M1}
T_{\rm nl}[\rho] \equiv 2\pi\int_0^\infty dr~r\tau_{\rm nl}(r),
\eeq
where the kinetic energy density is given by
\beq\label{Z5}
\tau_{\rm nl}(r) = \frac{3}{8}\rho(r)\int_0^\infty dk~k~J_0(kr)
\tilde{w} (k/2k_F(\rho(r)))\rho(k) - \frac{\pi}{4} \rho^2(r) +
\frac{1}{8\rho(r)}  \left |\frac{d\rho(r)}{dr} \right |^2~.
\eeq
Here, we have used the vW kinetic energy density of
Eq.~(\ref{C3}). If instead we use the form in Eq.~(\ref{C4}), we
obtain the kinetic energy density $\tau^{(1)}_{\rm nl}(r)$. Both
forms of the KE density yield the same kinetic energy as well as
the same set of self-consistent equations for the spatial
density, {\it viz.,} Eqs.~\eqref{SC8}--\eqref{SC11}.
%{\bf Please explain why this is so: \{}However, $\tau^{(1)}_{\rm nl}(r)$ turns out to be
%a more convenient form for the numerical calculations carried 
%out in the next subsection.{\bf \}}

In Table \ref{tableI}, the values of the exact KE, Eq.~\eqref{A3}, 
along with the results obtained from Eq.~\eqref{M1} using 
the exact densities as input are presented.
\begin{table}[ht] 
\centering      % used for centering table 
\begin{tabular}{c c c  c}  % centered columns (4 columns) 
\hline\hline                        %inserts double horizontal lines 
$N$ & $T_{\rm ex}$ & $T_{\rm nl}[\rho_{\rm ex}]$ & RPE \\ [0.5ex] % inserts table 
%heading 
\hline                    % inserts single horizontal line 
30    &  55   &   53.61     &   2.5\\
90   &  285 &    281.24    &   1.3\\
132 &    506   &     500.88    &    1.2\\ 
182 &  819 &  812.43 & 0.80\\ 
 420 & 2870 & 2857.8  & 0.42  \\[1ex]       % [1ex] adds vertical space 
\hline     %inserts single line 
\end{tabular} 
\caption{Comparison of the exact kinetic energies, $T_{\rm
ex}$, obtained 
from Eq.~\eqref{A3} and those obtained from Eq.~\eqref{M1},
$T_{\rm nl}[\rho_{\rm ex}]$, with 
the exact density used as input. 
The last column gives the relative percentage error (RPE) 
of the nonlocal functional result.
Energies are measured in units of $\hbar\omega$.} 
% title of Table 
\label{tableI}  % is used to refer this table in the text 
\end{table} 
The table illustrates that the nonlocal functional gives quite 
good results even for relatively low particle numbers,
and  by $N=420$, the relative percentage error (RPE) is 
already below $0.5 \%$.  

It would appear that the KE functional in Eq.~\eqref{M1} is
performing quite well, especially given the rather inhomogeneous 
nature of the harmonically-confined density. However, the
smallness of the RPEs in Table~\ref{tableI} is not a 
sufficiently stringent criterion for judging the accuracy of 
the functional. This is brought home by
the fact that  the TF ({\it i.e.,} LDA)  KE functional, 
\bea\label{A4}
T_{\rm TF}[\rho] &=& \int d^2r~\frac{\pi}{2}\rho^2(r)\nonumber \\
&\equiv& \int d^2r~\tau_{\rm TF}(r)~,
\eea
yields the {\it exact} KE when the exact density is used as input; that is, $T_{\rm TF}[\rho_{\rm ex}] = T_{\rm ex}$ 
for any number of closed shells~\cite{brack_vanzyl}. This
surprising result is a special property of the
harmonically-confined system in two-dimensions. 
%Consequently,  $T_{\rm TF}[\rho_{\rm ex}]$ is also exact in the extreme vW limit.  
Of course, $T_{\rm TF}[\rho]$ is also exact in the uniform gas limit.
Thus, as far as the KE is concerned, the absolutely crudest 
approximation for the KE functional for the 2D HO system
out-performs our nonlocal functional. The message to be
taken from this observation is that to test the quality of any 
proposed KE functional, one must go beyond simply investigating the 
global value it returns when available exact densities are used 
as input. Indeed, notwithstanding the results obtained using 
the TF functional for the 2D HO potential, the LDA KE functional is
certainly not exact.
%{\bf I'm more or less happy with everything up to this
%point.}\hfil \break

One might imagine that a better validation of the quality of 
$T_{\rm nl}[\rho]$ is provided by a point-wise comparison of 
$\tau_{\rm nl}(r)$ with the exact KE density.
\begin{figure}[t]
\centering \scalebox{0.8}
{\includegraphics{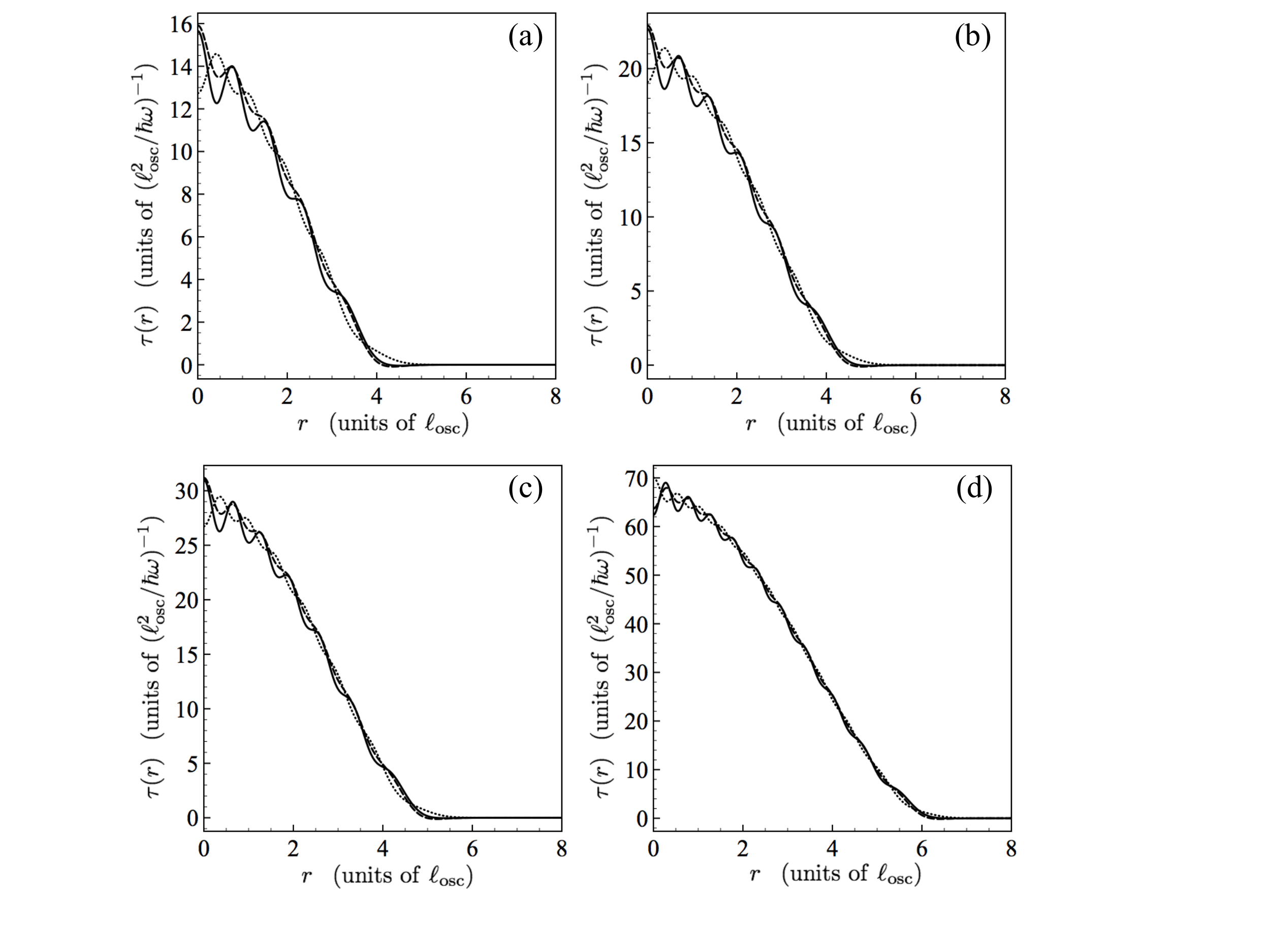}}
\caption{As in Fig.~\ref{fig1} but now the solid curves correspond to $\tau^{(1)}_{\rm nl}(r)=\tau^{(1)}_{\rm nl}(\rho_{\rm ex}(r)) $.}
\label{fig2}
\end{figure}
In Figs.~\ref{fig1} and \ref{fig2}, we present a comparison of 
the KE densities $\tau_{\rm nl}$ and $\tau^{(1)}_{\rm nl}$ with
$\rho_{\rm ex}(r)$ used as input (solid curves) with
Eq.~\eqref{keden} (dotted curves) and 
Eq.~\eqref{keden1} (dashed curves), respectively,
for a variety of particle numbers.  Focusing first on Fig.~\ref{fig1}, 
we note that the shell oscillations in $\tau_{\rm nl}(\rho_{\rm ex}(r))$
are reduced in amplitude when compared to the exact 
KE densities, and tend to be in phase with the oscillations
of $\tau^{(1)}_{\rm ex}(r)$ in the bulk.  However, for large-$r$, 
$\tau_{\rm nl}(\rho_{\rm ex}(r))$ begins to follow
$\tau_{\rm ex}(r)$ as it falls to zero from above for $r\to\infty$, which is expected in view
of Eq.~\eqref{keden_large}.  Figure~\ref{fig2} provides a
comparison of $\tau^{(1)}_{\rm nl}(\rho_{\rm ex}(r))$ with the 
exact KE densities. The feature which stands out most
dramatically is the enhanced shell oscillations in
$\tau^{(1)}_{\rm nl}(\rho_{\rm ex}(r))$.  Although
$\tau^{(1)}_{\rm nl}(\rho_{\rm ex}(r))$ closely matches the peaks 
of $\tau^{(1)}_{\rm ex}(r)$ in the bulk, it overshoots
the valleys by a large margin (see, {\it e.g.,} panel (a) in 
Fig.~\ref{fig2}).  Moreover, 
in the low-density tail region, $\tau^{(1)}_{\rm nl}(\rho_{\rm ex}(r))$
{\em continues } to follow $\tau^{(1)}_{\rm ex}(r)$ as it dips below zero, and then rises to zero from
below as  $r\to\infty$ (again, this is expected given the 
negative vW contribution which dominates 
$\tau^{(1)}_{\rm nl}(\rho_{\rm ex}(r))$ as $r\to\infty$).  We 
also observe that the shell oscillations in all of the KE 
densities in Figs.~\ref{fig1} and~\ref{fig2} become less 
pronounced as the particle number increases.  This behaviour 
may be understood by noting that~\cite{vanzyl2003}
\beq\label{TFlimit1}
\lim_{N\to \infty} \tau_{\rm ex}(r)= \lim_{N\to \infty} \tau^{(1)}_{\rm ex}(r)=\lim_{N\to \infty}  \tau^{(2)}_{\rm ex}(r)=\frac{1}{2\pi}
\left(M+\frac{3}{2}-\frac{1}{2}r^2\right)^2~,
\eeq
and
\beq\label{TFlimit2}
\lim_{N\to \infty} \rho_{\rm ex}(r) = \frac{1}{\pi}
\left(M+\frac{3}{2}-\frac{1}{2}r^2\right)~,
\eeq
which are the TF forms for the KE and spatial densities, respectively, with {\em no shell structure} present.
Moreover, it can also be shown that
\beq\label{TFlimit3}
\lim_{N\to \infty}  \tau_{\rm nl}(r)= \lim_{N\to \infty}  \tau^{(1)}_{\rm nl}(r) = \frac{1}{2\pi}
\left(M+\frac{3}{2}-\frac{1}{2}r^2\right)^2~.
\eeq

Although there is reasonable agreement between $\tau_{\rm nl}$
and $\tau^{(1)}_{\rm nl}$ and the exact KE densities, the TF
approximation is once again far superior in this regard.
The TF KE density, $\tau_{\rm TF}(r)=\pi\rho^2(r)/2$, with the 
exact density as input, almost perfectly reproduces the shell 
oscillations of the exact KE density~\cite{brack_vanzyl}, $\tau_{\rm ex}^{(1)}(r)$,
({\em for an arbitrary number of closed shells}), except near 
the tail region (see inset to Fig.~1, panel (a)).
Therefore, not only does the TF KE functional yield the exact 
KE for the exact density, it also provides an outstanding 
representation of the exact KE density. By comparision, the ADA
KE densities perform rather poorly. 

However, a true measure of the quality of a functional is the
accuracy of the density it yields on minimization. Here the TF
functional has serious shortcomings in that the minimizing
density is non-analytic at the edge where the density goes to
zero~\cite{brack_bhaduri}. With this density, the TF KE deviates
significantly from the exact value. We now turn to a test of the
nonlocal KE functional using densities determined 
self-consistently and examine how the resulting energy, and 
density profiles, compare with the exact results. 
%%%%%%%%%%%%%%%%%%%%%%%%%%%%%%%%%% SELF CONSISTENT SOLUTION %%%%%%%%%%%%%%%%%%%%%%%%%%
\subsection{Tests using self-consistent densities}\label{sc}
Even for the ideal gas~\cite{note_ideal}, the TFvW-like theory 
needs to be solved self-consistently in order to obtain the 
ground state density, along with any other associated equilibrium 
properties.  For the circular symmetry of the HO potential,
the closed set of equations, 
Eqs.~\eqref{SC8}--\eqref{SC11}, depend only on the radial 
variable, {\it viz.,}
\beq\label{B1}
-\frac{1}{2} \left(\frac{d^2}{dr^2} + \frac{1}{r}\frac{d}{dr}\right) \psi(r) + v_{\rm eff}(r) \psi(r) = \mu \psi(r)~,
\eeq
with
\beq\label{B2}
v_{\rm eff}(r) = -\frac{\pi}{2} \psi^2(r) + \phi(r) + \frac{1}{2}r^2~,
\eeq
\begin{eqnarray}\label{B3}
\phi(r) &=& \frac{3\pi}{4} \int_0^\infty{d k} \int_0^\infty dr_1
\left[\Omega\left(\frac{k}{2k_F(r)}\right)
+\tilde{w}\left(\frac{k}{2k_F(r_1)}\right)  \right]kr
_1J_0(kr)J_0(kr_1) \rho(r_1)~. 
\end{eqnarray}
$\Omega(\eta)$ is defined in Eq.~\eqref{Omega}, and we recall that 
$k_F(r) = \sqrt{2\pi\rho(r)}$ with $\psi(r) = \sqrt{\rho(r)}$.
The normalization of the density now reads
\beq\label{B5}
N(\mu) = 2\pi\int_0^\infty dr~r |\psi(r)|^2~.
\eeq
Self-consistent solutions to the above equations have been
obtained using the discrete Hankel transform method outlined 
in References~\cite{dht,arfken}.

Following the analysis of the last subsection, we present in 
Table \ref{table2} the results for the kinetic energies obtained
from  Eq.~\eqref{M1}  (or equivalently, Eq.~\eqref{weight6}), but 
now with the 
self-consistent density, $\rho_{\rm sc}(r)$, used as input.
Table \ref{table2} illustrates that the agreement between the exact KE, and the one generated from $T_{\rm nl}[\rho_{\rm sc}]$ is excellent.  
In fact, the KE is significantly better than the values obtained 
from $T_{\rm nl}[\rho_{\rm ex}]$ in Table~\ref{tableI}.  
This is somewhat unexpected since one typically finds 
$T_{\rm nl}[\rho_{\rm sc}]$ to be considerably worse than 
$T_{\rm nl}[\rho_{\rm ex}]$
(see, {\it e.g.,} Refs.~\cite{comb97,chacon85}).

\begin{table}[ht] 
\centering      % used for centering table 
\begin{tabular}{c c c  c}  % centered columns (4 columns) 
\hline\hline                        %inserts double horizontal lines 
$N$ & $T_{\rm ex}$ & $T_{\rm nl}[\rho_{\rm sc}]$ & RPE \\ [0.5ex] % inserts table 
%heading 
\hline                    % inserts single horizontal line 
30    &  55   &   54.28     &   1.3\\
90   &  285 &    283.16    &   0.64\\
132 &    506   &     503.60    &    0.47\\ 
182 &  819 &  816.11 & 0.35\\ 
 420 & 2870 & 2866.48  & 0.12 \\[1ex]       % [1ex] adds vertical space 
\hline     %inserts single line 
\end{tabular} 
\caption{Comparison of the exact kinetic energy  $T_{\rm ex}$ in 
Eq.~\eqref{A3} with the kinetic energy obtained from the
nonlocal functional $T_{\rm nl}[\rho_{\rm sc}]$ 
in Eq.~\eqref{M1}, evaluated using the self-consistent density $\rho_{\rm sc}(r)$.  The last column 
gives the relative percentage error (RPE) between 
$T_{\rm ex}$ and $T_{\rm nl}[\rho_{\rm sc}]$.
Energies are measured in units of $\hbar\omega$.} 
% title of Table 
\label{table2}  % is used to refer this table in the text 
\end{table} 
\begin{figure}[ht]
\centering \scalebox{0.7}
{\includegraphics{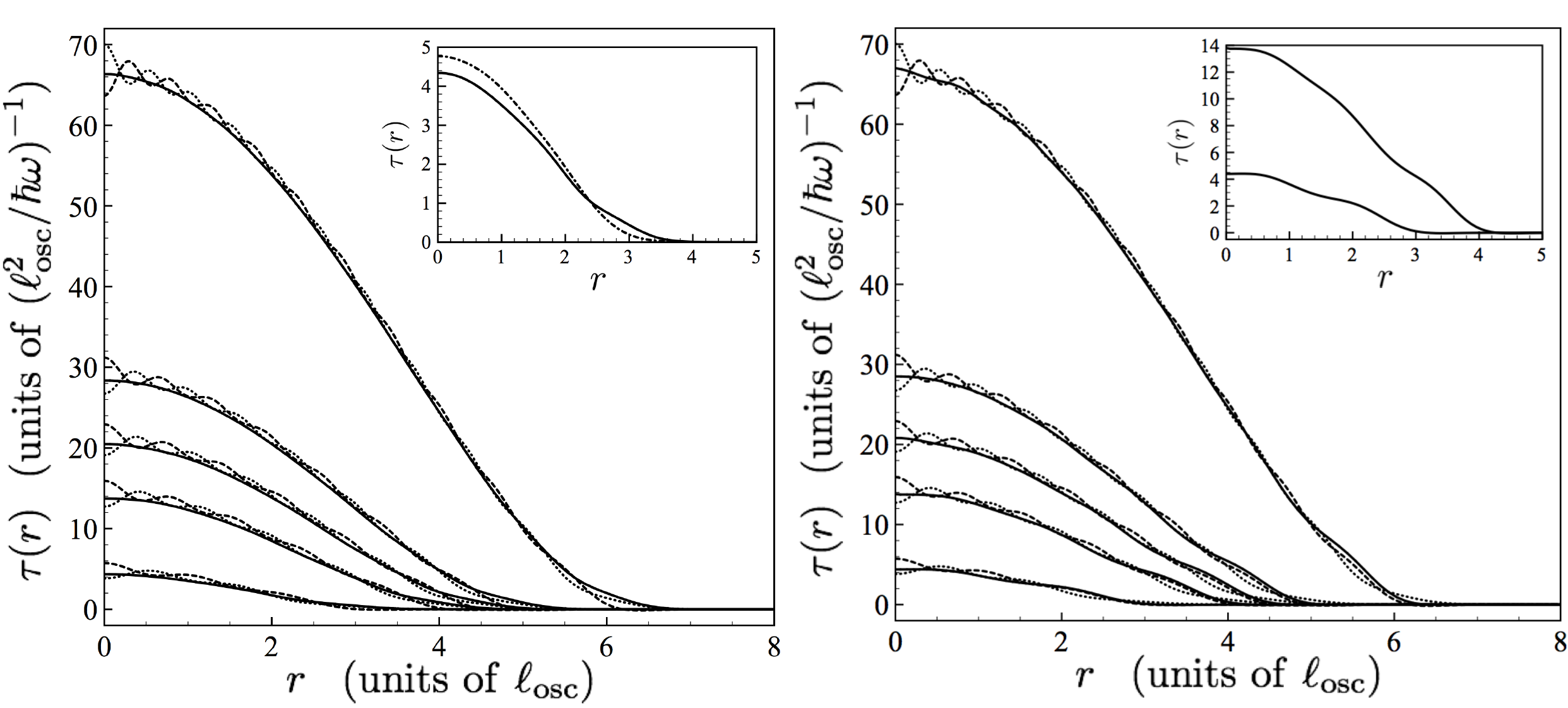}}
\caption{Left panel: The kinetic energy densities, $\tau_{\rm ex}(r)$ (dotted curve), $\tau^{(1)}_{\rm ex}(r)$ (dashed curve) and $\tau_{\rm nl}(r) = \tau_{\rm nl}(\rho_{\rm sc}(r))$  (solid curve), 
from lowest to highest curves, for $N=30, 90,132,182,420$ particles. Right panel:  As in the left panel, but
now for $\tau^{(1)}_{\rm nl}(r) = \tau^{(1)}_{\rm nl}(\rho_{\rm sc}(r))$  (solid curve).
The inset to the left panel
shows $\tau_{\rm nl}(r)$ and $\tau^{(2)}_{\rm ex}(r)$ (dot-dashed curve) for $N=30$ particles.
The inset to the right panel
shows  $\tau^{(1)}_{\rm nl}(r)$ for $N=30,90$ particles, highlighting the oscillatory structure present in the self-consistent KE density, $\tau^{(1)}_{\rm nl}(r)$.  The axes of the figure
insets are scaled as in the main figures.}
\label{fig3}
\end{figure}

Figure \ref{fig3} displays two panels.  On the left, the 
self-consistent KE densities $\tau_{\rm nl}(r)$ obtained 
from Eq.~\eqref{Z5}
(solid curves) are compared to the exact KE densities.  
The right panel shows a similar comparison for $\tau^{(1)}_{\rm
ex}(r)$.
Let us first focus on the left panel in Figure~\ref{fig3}.  
We see that $\tau_{\rm nl}(r) = \tau_{\rm nl}(\rho_{\rm sc}(r))$ 
exhibits a weak oscillatory structure which we will refer to as
`shell-like'. This structure is a consequence of the nonlocal
nature of the KE functional but should not be associated with
the shell structure arising from the occupancy of multiple
orbitals. In addition, $\tau_{\rm nl}(r)$ 
appears to be close to $\tau^{(2)}_{\rm ex}(r)$, the 
average of $\tau_{\rm ex}(r)$ and
$\tau^{(1)}_{\rm ex}(r)$.  The figure inset to the left panel 
shows, however, that $\tau_{\rm nl}(r)$ differs significantly
from  $\tau^{(2)}_{\rm ex}(r)$ (dot-dashed curve in the inset)
when $N$ is small, both within the bulk and in the low
density regions approaching the classical turning point.  
%The figure inset reveals that there are shell effects present, 
%but they are considerably muted.  
In the main figure, we once again observe that
$\tau_{\rm nl}(r)$ closely follows $\tau_{\rm ex}(r)$ for large-$r$, similar to what was found in Figure~\ref{fig1}.
Overall, aside from the weak oscillations, $\tau_{\rm nl}(r)$ is a reasonable representation of the exact
KE densities, particularly in the tail region, where it follows $\tau_{\rm ex}(r)$.

Moving on to the right panel in Fig.~\ref{fig3}, we observe that 
the primary difference between the self-consistent $\tau_{\rm nl}(r)$
and $\tau^{(1)}_{\rm nl}(r)$ is the presence of enhanced
shell-like oscillations in the latter (akin to what was found 
in Figure~\ref{fig2}). This structure is revealed more clearly 
in the inset to the right panel where we display the self-consistent
$\tau^{(1)}_{\rm nl}(r)$ for $N=30,90$ particles.
The other noteworthy 
difference between the two self-consistent KE densities in the 
left and right panels is that in the tail region, 
$\tau^{(1)}_{\rm nl}(r)$ nicely follows $\tau^{(1)}_{\rm ex}(r)$,
similar to what was found in Figure~\ref{fig2}.  Again, the 
large-$r$ behaviour of either self-consistent KE
density can be understood from the form of the vW terms being
used in $\tau_{\rm nl}(r)$ and $\tau^{(1)}_{\rm nl}(r)$, which
dominate at low densities.

\begin{figure}[H]
\centering \scalebox{0.27}
{\includegraphics{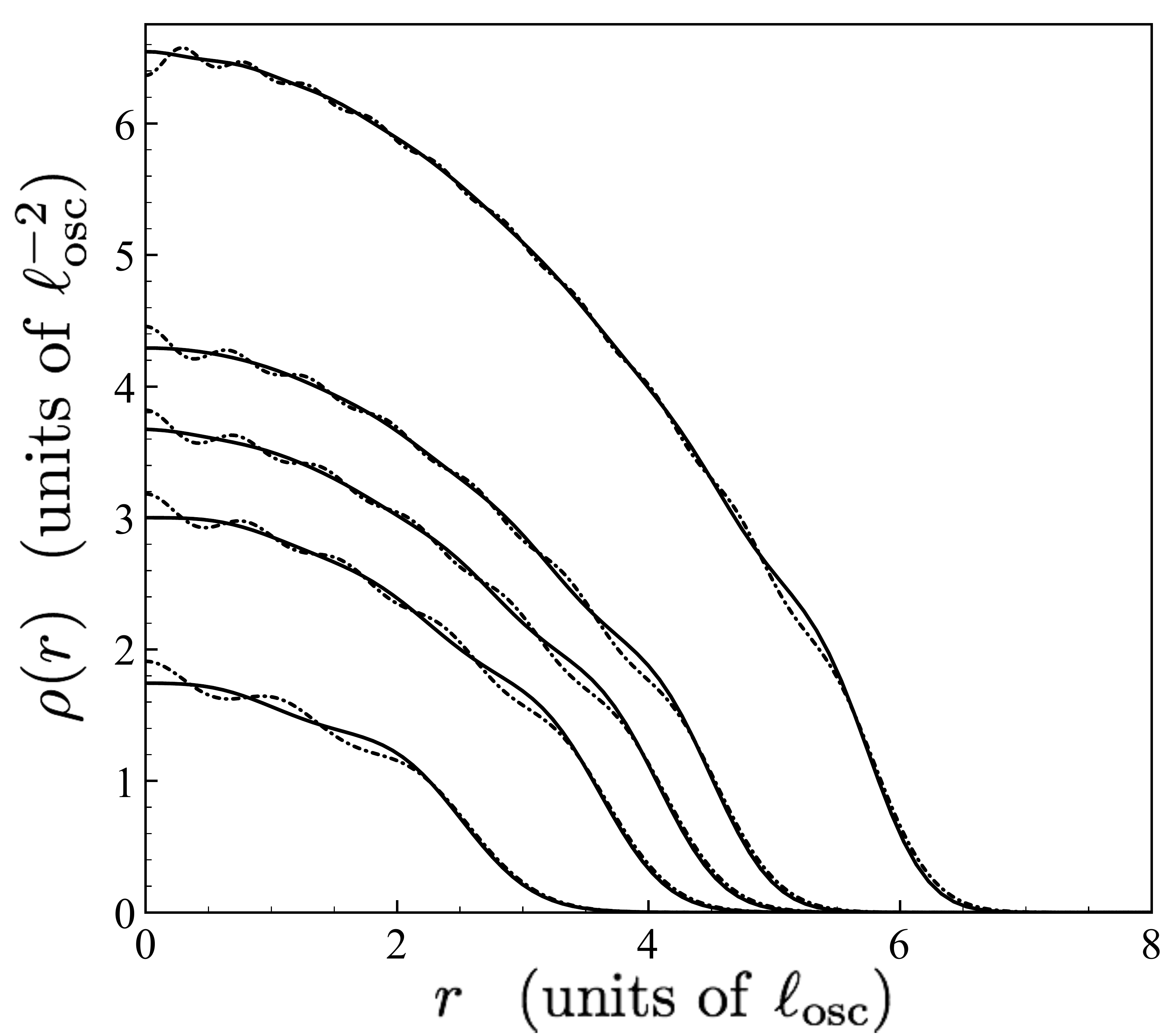}}
\caption{The exact (dot-dashed curves), and TFvW-like self-consistent  spatial densities (solid curves), from lowest to highest curves, for $N=30, 90,132,182,420$ particles.}
\label{fig4}
\end{figure}

In Fig.~\ref{fig4}, we compare the exact (dot-dashed curves) and self-consistent spatial densities (solid curves) for the particle numbers in Table \ref{table2}.  The self-consistent spatial densities also
display oscillatory structure, but these
%although they are significantly 
%diminished compared to the exact densities, especially as one 
%moves up in particle number. Note that the oscillations in the 
%self-consistent densities 
do not match up with the oscillations in the exact spatial 
densities, which emphasizes the fact that the oscillations have
different origins.

\begin{figure}[H]
\centering \scalebox{0.27}
{\includegraphics{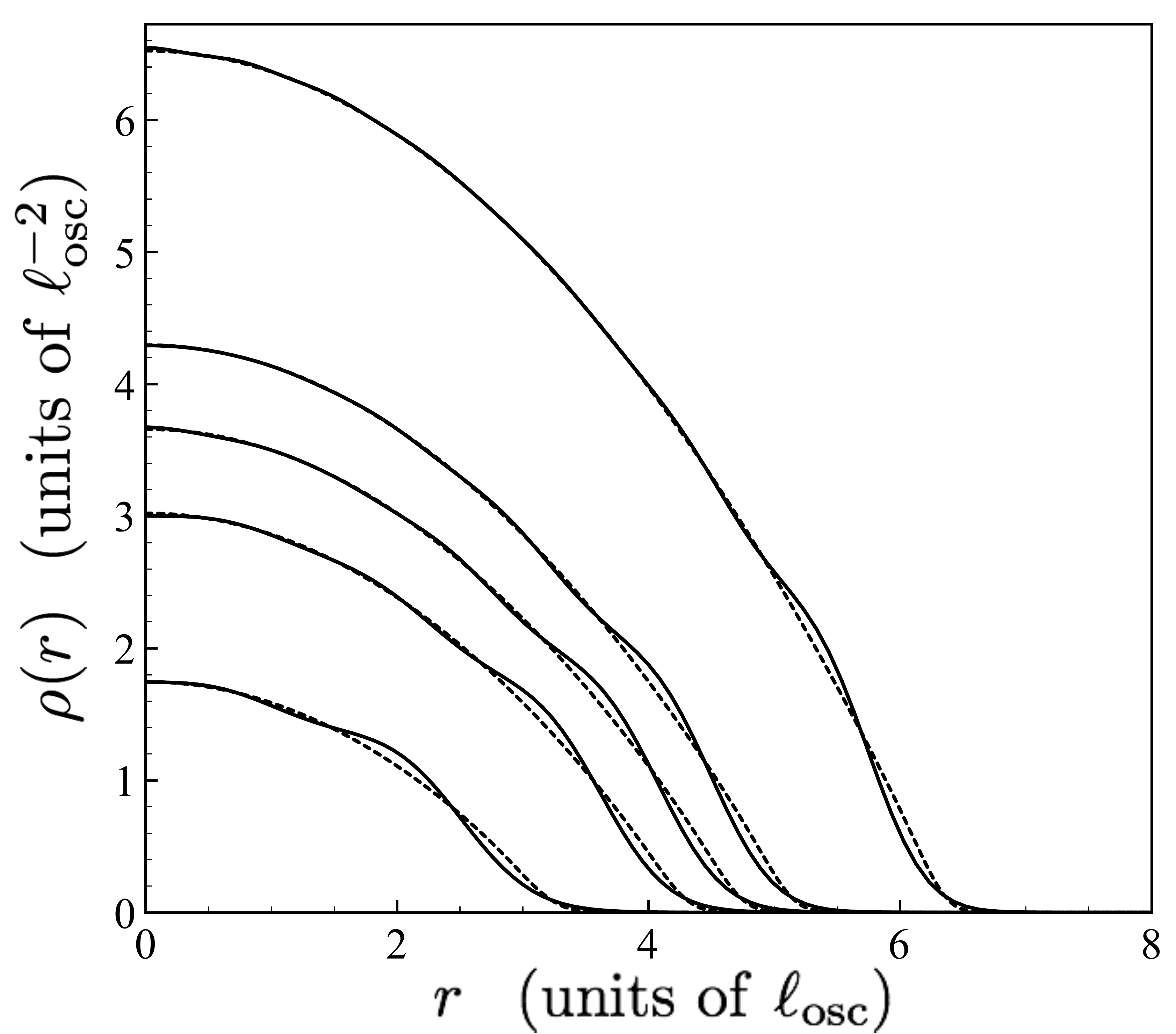}}
\caption{The self-consistent spatial densities from the nonlocal TFvW-like theory (solid curves) and TFvW theory (dashed curves) for $N=30,90,132,182,420$ particles.
Note that the TFvW densities are obtained by using an optimal vW coefficient for each particle number
($\lambda_{\rm vW} = 0.0568$, 0.0503, 0.0484, 
0.0468, 0.0433 for $N=30, 90, 132, 182, 420$ particles, 
respectively). }
\label{fig6}
\end{figure}

%the approximate treatment of two-point correlations
%in the nonlocal KE functional.  
More importantly, it is very encouraging to see that the 
self-consistent spatial densities accurately reproduce the 
shape of the exact densities as one moves out to the edge 
of the cloud.  For example, the lowest curve in Fig.~\ref{fig4}, 
corresponding to $N=30$ particles, shows excellent agreement 
with the exact density for $r\gtrsim 2.5$.  The fact that 
the TFvW-like self-consistent densities yield the correct behaviour  
near the edge of the distribution is a significant result given that it is obtained with no adjustable parameters.
To further highlight the quality of the self-consistent TFvW-like theory, we present in Fig.~\ref{fig6}
a comparison of the TFvW and TFvW-like spatial densities for $N=30,90,132,182,420$.  
It is clear that the optimal TFvW densities (dashed curves) exhibit too sharp
a decay into the classically forbidden region, as compared to the TFvW-like self-consistent densities (solid curves).  In fact, we have examined the TFvW spatial densities for
a range of values, $0 < \lambda_{\rm vW} \leq 1$, and have found that 
they cannot provide the correct density profile of the exact densities near the edge of the system.  
In contrast, the nonlocal TFvW-like theory
correctly captures the surface profile of the exact density (see Fig.~\ref{fig4}) with no adjustable parameters.

\begin{figure}[H]
\centering \scalebox{0.27}
{\includegraphics{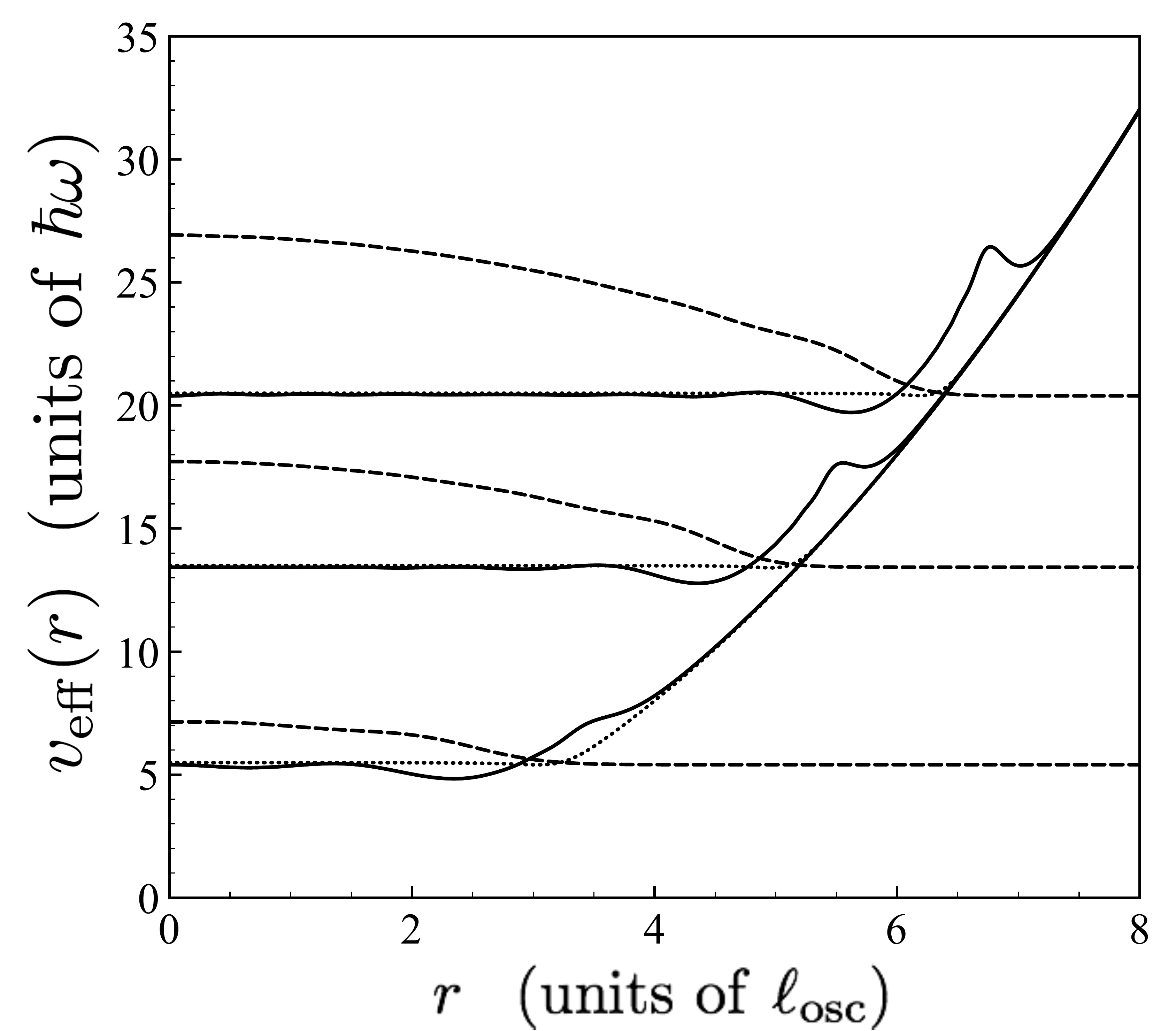}}
\caption{The self-consistent effective potentials for $N=30,182,420$ particles corresponding to the TFvW (dotted curves) and TFvW-like (solid curves).  The dashed curves are the self-consistent
TFvW-like densities, which have been overlaid to illustrate the way in which $v_{\rm eff}(r)$ determines the density distribution.}
\label{figbw5and6}
\end{figure}

In order to gain some insight into the differences between the TFvW and TFvW-like spatial densities, 
we display in Fig.~\ref{figbw5and6} the self-consistent effective potentials, $v_{\rm eff}(r)$, from the
two theories.   Our first observation is that the oscillatory structure of the TFvW-like $v_{\rm eff}(r)$ manifests itself in the self-consistent TFvW-like spatial density (see, {\it e.g.,} the lowest set of curves
in Fig.~\ref{figbw5and6}, where the TFvW-like spatial densities have been overlaid as the dashed-curves).  This of course is not surprising, since the effective potential determines the spatial density profile of the system through Equation~\eqref{B1}.  
In addition, both the TFvW and TFvW-like $v_{\rm eff}(r)$ are in good agreement near the centre of the trap, and with increasing particle number, the TFvW-like effective potential
closely follows the TFvW curve in the bulk; this agreement explains the similarities between the TFvW and TFvW-like spatial densities displayed in Fig.~\ref{fig6} for large particle numbers.   
However, as we approach the edge of the system ({\it e.g.,} for the lowest curves in Fig.~\ref{figbw5and6}, $r\gtrsim 2$)
the differences in the structure of the effective potentials in the low-density regime are quite dramatic.  Specifically, the TFvW-like effective potential develops a local minimum near the edge
of the cloud, whereas for the same coordinate range, the TFvW potentials remain comparatively flat.  In view of the results presented in Fig.~\ref{fig4}, it is clear that the behaviour of the TFvW-like effective
potential near the surface of the system is a more accurate representation of the exact effective potential. The other noteworthy feature in the TFvW-like effective potential is the ``knee'' developing in the very low-density regime, which becomes more prominent as the number of particles
is increased.  We have numerically determined that this knee-like
feature is a result of the {\em nonlocal} part of the effective potential, arising specifically from the $\tilde{w}$ term in $\phi(r)$ in Equation~\eqref{B3}.  While this knee feature is only prominent in the low density 
regime, its presence appears to be crucial for providing the correct behaviour of the TFvW-like spatial density in the classically forbidden region, $r\gg R_{\rm TF}$.  It would be of interest to examine
the {\em exact} effective potential obtained from the 2D HO KE density functional~\cite{march}, against the TFvW-like effective potentials shown in Figure~\ref{figbw5and6} obtained from the ADA nonlocal KE functional. 

\section{Closing Remarks}
We have applied the ADA in the construction of a nonlocal 
KE functional, and have used it to formulate a TFvW-like theory 
for the ground state properties of a 2D inhomogenous Fermi gas.  
One of our central findings is that the 2D ADA nonlocal KE 
functional does not admit additional parameters in its
definition, in contrast to the situation in 1D and 3D, where there is considerable freedom in specifying the form of the functional. 
In addition, the ADA naturally leads to a vW term in 2D, which is consistent with what is found in other dimensions. 
Although such a gradient correction
cannot be justified in 2D on the basis of a systematic gradient
expansion, it is nevertheless an important component of the
nonlocal kinetic energy functional, particularly in the low-density region, where the decay into the classically forbidden region is smooth.

A commonly-used procedure for testing the efficacy of a KE
functional is to investigate its ability to generate accurate energies when using the exact density 
for some model situation. We point out that the results of such
a test for a 2D harmonically confined, ideal Fermi gas can be
misleading, to wit, in the case of 2D HO confinement, even the crudest, local TF kinetic energy
functional, yields superlative results. However, a true measure
of the quality of a functional can only be ascertained by examining the nature of the density profiles it produces upon
a functional minimization
with respect to the density. In this regard, the TF
functional demonstratably fails. On the other hand, the fully
self-consistent DFT calculations we have performed using the ADA
kinetic energy functional yield very good results for the
total energy of the harmonically-confined model system. In 
addition, based on a comparison with exact results, we find 
that the TFvW-like theory provides a surprisingly good
description of the density in the low-density regime.

We have also compared our self-consistent calculations with
the results of an earlier 2D TFvW theory in which the vW
coefficient is optimized in order to yield the correct total
energy. This comparison shows that the local nature of the TFvW
KE functional, Eq.~\eqref{vW2D}, is the reason behind the poor
description of the surface density profile. 
Nevertheless, the reasonably good agreement with exact results,
along with its simple form, and
exceedingly easy numerical implementation, 
suggest that the 2D TFvW is still a 
useful tool for the description of inhomogeneous 2D systems, provided one is interested
in properties that are 
relatively insensitive to the local details of the equilibrium spatial density
({\it e.g.,} total energies, and collective excitations~\cite{vanzyl1,vanzyl2,vanzyl3,vanzyl4}). 

Finally, we believe that this paper fills a gap in the literature dealing with the DFT of 2D non-uniform Fermi systems.    In particular, where 
standard linear response and semiclassical expansion techniqes in 2D fail to produce gradient corrections associated with spatial inhomogenieties, the ADA naturally allows for the inclusion of beyond LDA physics to the KE functional in a dimensionally independent way.   
More importantly, the 2D KE functional
developed within the ADA
has {\em no free parameters}, so it may be scrutinized in other systems without the possibility of any ``fine tuning''.
Our hope is that the nonlocal functional we have presented will 
stimulate further work toward the development of more accurate KE functionals in low-dimensional Fermi systems.  For example, an interesting extension of this work would be to apply the generalization of the ADA 
developed in
Ref.~\cite{garcia96} to 2D, and examine how the resulting functional improves the determination of the global, and local  properties ({\it e.g.,} shell structure) of the system.

\acknowledgements
This work was supported by the Natural Sciences and Engineering Research Council of Canada (NSERC), and the National Research Foundation and  Ministry of Education, Singapore.


\begin{thebibliography}{99}
\bibitem{note_nl}
Only a truly nonlocal KE functional can satisfy the exact linear response condition, Eq.~\eqref{W4}, in two-dimensions.
\bibitem{note_ideal}
We are limiting our self-consistent calculation to the noninteracting case, so that we can make comparisons with the available exact results.   The inclusion of interactions 
will not alter the general conclusions drawn in this paper.  Moreover, eliminating interactions allows us to attribute the oscillatory structure found in the spatial and KE densities directly to the nonlocal KE functional.
\end{thebibliography}
\end{document}